\documentclass[aip,reprint]{revtex4-1}
\usepackage{graphicx}
\usepackage{appendix}
\usepackage{braket}
\usepackage{rotating}
\usepackage{amssymb}
\usepackage{amsmath}
\usepackage{txfonts}
\usepackage{bm}
\usepackage{color}
\usepackage{multirow}
\usepackage{booktabs}
\usepackage{dcolumn}

\begin{document}

\title{Thermal mean-field theories}

\author{Pinhao Gu}
\affiliation{Department of Chemistry, University of Illinois at Urbana-Champaign, Urbana, Illinois 61801, USA}

\author{So Hirata}
\email{sohirata@illinois.edu}
\affiliation{Department of Chemistry, University of Illinois at Urbana-Champaign, Urbana, Illinois 61801, USA}

\date{\today}

\begin{abstract}
Several closely related {\it ab initio} thermal mean-field theories for fermions, both well-established and new ones, are compared 
with one another at the formalism level and numerically. The theories considered are Fermi--Dirac theory, thermal Hartree--Fock (HF) theory, two modifications 
of the thermal single-determinant approximation of Kaplan and Argyres, and first-order finite-temperature many-body perturbation theory based on zero-temperature or thermal
HF reference. The thermal full-configuration-interaction theory is used as the benchmark.
\end{abstract}

\maketitle 

\section{Introduction}

The zero-temperature Hartree--Fock (HF) theory is the cornerstone of all {\it ab initio} electronic structure theories.\cite{szabo,shavitt} It is stipulated as 
the minimization of the expectation value of the exact interacting Hamiltonian in a Slater determinant by varying orbitals, while maintaining their orthonormality.
This leads to Brillouin's condition,\cite{szabo} which requires a certain relation be satisfied by a pair of orbitals, one occupied and the other unoccupied, in the Slater determinant.
Lagrange's undetermined multipliers $\epsilon_{pq}$ for the orthonormality constraint, when diagonalized, can be interpreted as the approximate 
one-electron energies according to Koopmans' theorem.\cite{szabo} They provide a compelling theoretical basis for the one-electron picture of 
chemistry including energy bands in solids.

Thermal HF theory\cite{Husimi1940,Mermin63} is formulated
in a surprisingly different manner while correctly reducing to the zero-temperature counterpart as temperature ($T$) goes to zero. The $\epsilon_{pq}$, which eventually 
become the thermal HF ``orbital energies,'' 
are variational parameters defining an auxiliary non-interacting Hamiltonian. Thermodynamic functions for the exact interacting Hamiltonian 
are then formulated with a one-particle density matrix of this non-interacting Hamiltonian. 
The grand potential is minimized by varying $\epsilon_{pq}$. This theory does not seem to have a single, consistent grand partition function. Nor does it lead to 
a condition that reduces to Brillouin's condition (whose ``occupied'' and ``unoccupied'' designations seem incongruous to thermodynamics in the first place).

Perhaps because of this disparity, there have been confusions about thermal HF theory. 
Kirzhnits\cite{Kirzhnits} stated that the theory does not obey fundamental thermodynamic relations, which was subsequently
overturned by Argyres {\it et al.}\cite{Argyres1974} 
Kaplan and Argyres\cite{Kaplan1975} furthermore introduced 
an alternative (or perhaps even antithetical) thermal HF-like ansatz they called the thermal single-determinant approximation (TSDA), which is derivable from a single, consistent grand partition function. 

The physical meaning of thermal HF orbital energies is still obscure,\cite{Pain} 
even though they reduce to the zero-temperature HF orbital energies with well-established physical meaning. They are 
frequently and uncritically invoked to rationalize electronic metal-insulator transitions, etc.\cite{Hermes2015}
The fact that they vary with temperature also seems to defy their interpretation as quantized state energies of some kind. 

In the last couple of years, one of the authors with coauthors 
established finite-temperature many-body perturbation theories for electrons\cite{HirataJha,HirataJhaJCP2020,Hirata2021} and anharmonic phonons,\cite{Qin2023} 
which expand all thermodynamic functions in consistent power series. In one of these articles,\cite{Hirata2021} it was pointed out that the thermal HF grand potential expression 
does not agree with the first-order grand potential expression. Nor does the thermal HF internal energy expression coincide with the first-order internal energy expression. 
This is unlike the zero-temperature case, where 
the first-order M{\o}ller--Plesset perturbation energy is identified as the HF energy, with the second order being the leading order in describing electron correlation.\cite{szabo,shavitt} 

In this article, we rigorously and pedagogically examine 
the formalisms of several thermal {\it ab initio} mean-field theories, both well-established and new ones, 
such as Fermi--Dirac theory, thermal HF theory, two modifications of the TSDA of Kaplan and Argyres,\cite{Kaplan1975} 
and first-order finite-temperature many-body perturbation theory (MBPT). (Here, ``mean-field theories'' merely refer to the ones that do not account for electron correlation.)
They are also compared numerically with one another as well as with the thermal full-configuration-interaction (FCI) theory\cite{Kou} as the exact treatment within a 
basis set. Particular emphases are placed on 
their mutual relationships and numerical performance as well as on whether they satisfy thermodynamic relations. 

\section{Thermodynamic relations}

Let us succinctly summarize the definitions of exact thermodynamic functions and their mutual relationships.

The grand partition function $\Xi$ for interacting particles in the grand canonical ensemble is defined by
\begin{eqnarray}
\Xi \equiv \sum_I e^{-\beta (E_I - \mu N_I)} , \label{Exact:Xi} 
\end{eqnarray}
where $\beta = (k_\text{B}T)^{-1}$ is the inverse temperature,
the sum is taken over all states with any number of particles, $E_I$ is the energy of the $I$th state,
$N_I$ is the number of particles in the same state, and $\mu$ is the chemical potential. 
The thermal population (or diagonal density matrix element) for the $I$th state is then given by
\begin{eqnarray}
\rho_I \equiv \frac{e^{-\beta (E_I - \mu N_I)}}{\Xi}, \label{Exact:P} 
\end{eqnarray}
which sums to unity. 

Thermodynamic functions such as the grand potential $\Omega$, internal energy $U$, and entropy $S$
are derived from $\Xi$ or equivalently from $\rho_I$.
\begin{eqnarray}
\Omega &\equiv& -\frac{1}{\beta} \ln \Xi  \label{Exact:Omega1}  \\
&=& \sum_I \rho_I (E_I - \mu N_I) + \frac{1}{\beta} \sum_I \rho_I \ln \rho_I , \label{Exact:Omega2}  \\
U &\equiv&-\frac{\partial}{\partial \beta} \ln \Xi + \mu\bar N, \label{Exact:U1} \\ 
&=&  \sum_I \rho_I E_I, \label{Exact:U2} \\
\bar{N} &\equiv& \frac{1}{\beta} \frac{\partial}{\partial \mu}\ln \Xi \label{Exact:mu1}  \\
&=&  \sum_I \rho_I N_I, \label{Exact:mu2} \\
S &\equiv& \frac{U - \Omega -\mu\bar{N}}{T} \label{Exact:S1} \\
&=& -k_{\text{B}}  \sum_I \rho_I \ln \rho_I , \label{Exact:S2} 
\end{eqnarray}
where Eq.\ (\ref{Exact:mu1}) or (\ref{Exact:mu2}) is the condition determining $\mu$ for a fixed 
average number of particles $\bar{N}$ that maintains electroneutrality when the particle is electrically charged;\cite{Fisher,Dyson,HirataOhnishi} otherwise,
$\mu$ is arbitrary and $\bar{N}$ can vary.

These thermodynamic functions satisfy the relation,
\begin{eqnarray}
\Omega &=& U - \mu \bar{N} - TS,  \label{Exact:OmegaUmuS} 
\end{eqnarray}
as well as others such as 
\begin{eqnarray}
\bar{N} &=& -\frac{\partial\Omega}{\partial \mu}, \label{Exact:relation1}\\
S &=& -\frac{\partial \Omega}{\partial T}. \label{Exact:relation2}
\end{eqnarray}


In the following, 
we consider the case of electrons interacting through two-body potentials 
(with the repulsion energy of spatially-fixed nuclei set to zero) as a concrete example, but the conclusion should be valid for other fermions interacting through more complex potentials. 

\section{Fermi--Dirac theory\label{sec:FD}}

Fermi--Dirac theory is an exact treatment of non-interacting electrons in the grand canonical ensemble.
It therefore obeys all thermodynamic relations exactly. The Hamiltonian and number operator have the one-electron form,
\begin{eqnarray}
\hat{H}_0 &=& \sum_p^{\text{all}} \epsilon_p^{(0)} \hat{p}^\dagger \hat{p}, \label{H0} \\
\hat{N} &=& \sum_p^{\text{all}}  \hat{p}^\dagger \hat{p},
\end{eqnarray}
where letters $p$, $q$,  $r$, and $s$ are used for any spinorbital, either occupied or unoccupied (``all'') in any specific wave function,
and $\hat{p}^\dagger$ and $\hat{p}$ are electron creation and annihilation operators in the $p$th spinorbital with energy $\epsilon_p^{(0)}$. 
There is no special requirements on $\epsilon_p^{(0)}$ other than their reality, and it is a constant. The superscript `(0)'and subscript `0' are there 
because they define a zeroth-order (reference) theory for finite-temperature MBPT to be discussed in Sec.\ \ref{sec:PT}. 

Slater determinants $|I\rangle$ are the eigenfunctions of the corresponding Schr\"{o}dinger equation for any number of electrons.
For the $I$th state, the energy and number of electrons are given by
\begin{eqnarray}
E_I^{(0)} &=& \langle I | \hat{H}_0 | I \rangle = \sum_i^{\text{occ.}} \epsilon_i^{(0)}, \\
N_I &=&  \langle I | \hat{N} | I \rangle =\sum_i^{\text{occ.}} n_i,
\end{eqnarray}
where letters $i$, $j$, and $k$ are used for a spinorbital occupied (``occ.'') by an electron in the $I$th state ($n_i = 1$). 

The grand partition function then undergoes a drastic simplification into a sum-over-orbitals expression.
\begin{eqnarray}
\Xi^{(0)} &\equiv& \sum_{I} e^{-\beta( E_I^{(0)} - \mu^{(0)} N_I)} \nonumber \\
 &=& \prod_p \left( 1 + e^{-\beta(\epsilon_p^{(0)} - \mu^{(0)})} \right) \nonumber\\
 &=& \prod_p \frac{1}{f_p^+},
\end{eqnarray}
where the Fermi--Dirac distribution functions $f_p^\mp$ are given by
\begin{eqnarray}
f_p^- &=& \frac{1}{1+e^{\beta (\epsilon_{p}^{(0)} - \mu^{(0)})}}, \label{FD:minus} \\
f_p^+ &=&1 - f_p^- = \frac{e^{\beta (\epsilon_{p}^{(0)}-\mu^{(0)})}}{1+e^{\beta (\epsilon_{p}^{(0)}-\mu^{(0)})}}. \label{FD:plus}
\end{eqnarray}

The corresponding grand potential $\Omega^{(0)}$, internal energy $U^{(0)}$, and average number of electrons $\bar{N}$ are then
obtained from their definitions as
\begin{eqnarray}
\Omega^{(0)} &\equiv& -\frac{1}{\beta} \ln \Xi^{(0)} \label{FD:Omega} = \frac{1}{\beta} \sum_p \ln f_p^+ \\
&=& \sum_p \left( \epsilon_p^{(0)} -\mu^{(0)} \right) f_p^-  + \frac{1}{\beta} \sum_p \left( f_p^- \ln f_p^- + f_p^+ \ln f_p^+  \right), \nonumber\\ \label{FD:Omega2}\\
U^{(0)} &\equiv& -\frac{\partial}{\partial \beta} \ln \Xi^{(0)}+\mu^{(0)}\bar{N} \label{FD:U} =  \sum_p \epsilon_p^{(0)} f_p^-, \\
\bar{N} &\equiv& \frac{1}{\beta} \frac{\partial}{\partial \mu^{(0)}}\ln \Xi^{(0)} = \sum_p f_p^-, \label{FD:mu}
\end{eqnarray}
the last one being the condition determining $\mu^{(0)}$. They imply the entropy formula,
\begin{eqnarray}
S^{(0)} &=& -k_{\text{B}}  \left( f_p^- \ln f_p^- + f_p^+ \ln f_p^+  \right).
\end{eqnarray}

Together, these thermodynamic functions obey the thermodynamic relation,
\begin{eqnarray}
\Omega^{(0)} &\equiv& U^{(0)} - \mu^{(0)} \bar{N} - TS^{(0)},  \label{FD:S} 
\end{eqnarray}
and
\begin{eqnarray}
\bar{N} &=& -\frac{\partial\Omega^{(0)}}{\partial \mu^{(0)}}, \label{FD:relation1}\\
S^{(0)} &=& -\frac{\partial \Omega^{(0)}}{\partial T}. \label{FD:relation2}
\end{eqnarray}

Equations (\ref{FD:Omega}) and (\ref{FD:mu}) immediately imply Eq.\ (\ref{FD:relation1}). However, for future use, we explicitly show this, starting with Eq.\ (\ref{FD:Omega2}).
\begin{eqnarray}
- \frac{\partial \Omega^{(0)}}{\partial \mu^{(0)}} &=& \sum_p f_p^-  - \sum_p \left( \epsilon_p^{(0)} - \mu^{(0)} \right) \frac{\partial f_p^-}{\partial \mu^{(0)}}
\nonumber\\&&
 -\frac{1}{\beta} \sum_p \frac{\partial f_p^-}{\partial \mu^{(0)}} \ln f_p^- -\frac{1}{\beta} \sum_p \frac{\partial f_p^-}{\partial \mu^{(0)}}
 \nonumber\\&&
-\frac{1}{\beta} \sum_p \frac{\partial f_p^+}{\partial \mu^{(0)}} \ln f_p^+ -\frac{1}{\beta} \sum_p \frac{\partial f_p^+}{\partial \mu^{(0)}}
\nonumber\\&=&
\sum_p f_p^-  - \sum_p \left( \epsilon_p^{(0)} - \mu^{(0)} \right) \frac{\partial f_p^-}{\partial \mu^{(0)}}
\nonumber\\&&
 -\frac{1}{\beta} \sum_p \frac{\partial f_p^-}{\partial \mu^{(0)}}  \ln \frac{f_p^-}{f_p^+}
 \nonumber\\&=& \sum_p f_p^- 
 = \bar{N}, \label{FD:relation1explicit}
\end{eqnarray}
where we used
\begin{eqnarray}
\frac{\partial f_p^+}{\partial \mu^{(0)}} &=& - \frac{\partial f_p^-}{\partial \mu^{(0)}} , \\
\ln \frac{f_p^-}{f_p^+} &=& -\beta \left( \epsilon_p^{(0)} - \mu^{(0)} \right). \label{lnfoverf}
\end{eqnarray}
In this context, it would be wrong to write $\mu^{(0)}\bar{N}$ instead of $\sum_p \mu^{(0)} f_p^-$ in the definition of $\Omega^{(0)}$ [Eq.\ (\ref{FD:Omega2})] 
 because when a partial derivative with respect to $\mu^{(0)}$ is taken, it is $\beta$ that is being held fixed and not $\bar{N}$. 
An explicit evaluation of $\partial f_p^- / \partial \mu^{(0)}$ is never needed to prove the above identity, even though it is readily available as
\begin{eqnarray}
\frac{\partial f_p^\pm}{\partial \mu^{(0)}} &=& \mp \beta f_p^- f_p^+ .\label{mu_deriv}
\end{eqnarray}

Likewise, Eq.\ (\ref{FD:relation2}) is explicitly proven, starting with Eq.\ (\ref{FD:Omega2}).
\begin{eqnarray}
-\frac{\partial \Omega^{(0)}}{\partial T} &=& k_{\text{B}}\beta^2 \frac{\partial\Omega^{(0)}}{\partial \beta} 
\nonumber\\
&=& k_{\text{B}}\beta^2 \sum_p \left(\epsilon_p^{(0)} - \mu^{(0)}\right) \frac{\partial f_p^-}{\partial \beta} 
\nonumber\\&& 
- k_{\text{B}}  \left( f_p^- \ln f_p^- + f_p^+ \ln f_p^+  \right)
\nonumber\\
&& + k_{\text{B}}\beta \sum_p \left(\frac{\partial f_p^-}{\partial \beta} \ln f_p^- + \frac{\partial f_p^-}{\partial \beta} + \frac{\partial f_p^+}{\partial \beta} \ln f_p^+ + \frac{\partial f_p^+}{\partial \beta} \right) 
\nonumber\\
&=& k_{\text{B}}\beta^2 \sum_p \left(\epsilon_p^{(0)} - \mu^{(0)}\right) \frac{\partial f_p^-}{\partial \beta} 
\nonumber\\&&
- k_{\text{B}}  \left( f_p^- \ln f_p^- + f_p^+ \ln f_p^+  \right)
 + k_{\text{B}}\beta \sum_p \frac{\partial f_p^-}{\partial \beta} \ln \frac{f_p^-}{f_p^+} 
\nonumber\\
&=& - k_{\text{B}}  \left( f_p^- \ln f_p^- + f_p^+ \ln f_p^+  \right) = S^{(0)}, \label{FD:relation2explicit}
\end{eqnarray}
where an explicit evaluation of $\partial f_p^- / \partial \beta$ is again unnecessary, although it is available,
\begin{eqnarray}
\frac{\partial f_p^\pm}{\partial \beta} &=& \pm \left(\epsilon_p^{(0)} - \mu^{(0)}\right) f_p^- f_p^+ . \label{beta_deriv}
\end{eqnarray}

The thermal population of the $I$th state reduces to
\begin{eqnarray}
\rho_I^{(0)} &=& \frac{e^{-\beta (E_I^{(0)} -\mu^{(0)} N_I)}}{\sum_J e^{-\beta (E_J - \mu^{(0)} N_J)}}  \\
&=& \frac{\prod_i^{\text{occ.}} e^{-\beta(\epsilon_i ^{(0)}- \mu^{(0)})}}{\prod_p^{\text{all}} \left( 1 + e^{-\beta(\epsilon_p^{(0)} - \mu^{(0)})}\right)} \label{Boltzmannidentity} \\
&=& \prod_i^{\text{occ.}} f_i^- \prod_a^{\text{vir.}} f_a^+ ,
\end{eqnarray}
where ``vir.''\ stands for virtual (unoccupied) spinorbitals in the $I$th Slater determinant.
This simplified form of $\rho_I^{(0)}$ facilitates the evaluation of the density-matrix definitions
of $U^{(0)}$ and $\bar{N}$ [Eqs.\ (\ref{Exact:U2}) and (\ref{Exact:mu2})],
\begin{eqnarray}
U^{(0)} &=& \sum_I \rho_I^{(0)} E_I^{(0)} 
= \sum_p  \epsilon_p^{(0)} f_p^-, \\
\bar{N} &=& \sum_I \rho_I^{(0)} 
= \sum_p f_p^-,
\end{eqnarray}
where Boltzmann-sum identity I of Ref.\ \onlinecite{HirataJhaJCP2020} can be used in conjunction with Eq.\ (\ref{Boltzmannidentity}) to reach the final expressions.

Furthermore, this density matrix minimizes $\Omega^{(0)}$ because
\begin{eqnarray}
\frac{\partial \Omega^{(0)}}{\partial f_p^-} &=& \left( \epsilon_p^{(0)} -\mu^{(0)} \right) + \frac{1}{\beta} \ln \frac{f_p^-}{f_p^+}
= 0. \label{FD:variational}
\end{eqnarray}
Therefore, Fermi--Dirac theory is variational. 

\section{Thermal Hartree--Fock theory}

The derivation of thermal HF theory\cite{Husimi1940,Mermin63} by Mermin\cite{Mermin63} follows a rather different path; 
it does not start from a well-defined grand partition function given in terms of a single, consistent Hamiltonian. 

Instead, we define a {\it first} Hamiltonian of the one-electron form,
\begin{eqnarray}
\hat{H}_0 = \sum_p \epsilon_p^{\text{HF}} \hat{p}^\dagger \hat{p},
\end{eqnarray}
where spinorbitals labeled by $p$ are the ones that bring the Hermitian matrix of {\it variational parameters} $\bm{\epsilon}^\text{HF}$ into a diagonal form
with $\epsilon_p^\text{HF}$ being the $p$th eigenvalue.\cite{Mermin63}

Slater determinants made of these spinorbitals are the eigenfunctions of the first Hamiltonian $\hat{H}_0$ with eigenvalues,
\begin{eqnarray}
E_I^{\text{HF}1} &=& \langle I | \hat{H}_0| I \rangle = \sum_i^{\text{occ.}} \epsilon_i^{\text{HF}}, \\
N_I &=& \langle I | \hat{N}| I \rangle = \sum_i^{\text{occ.}} n_i,
\end{eqnarray}
where superscript `HF1' is there to  distinguish these state energies from the second definition of state energies to be introduced shortly. 

The corresponding thermal population undergoes the same drastic simplification as that of Fermi--Dirac theory and becomes
\begin{eqnarray}
\rho_I^{\text{HF}1} &=& \frac{e^{-\beta (E_I^{\text{HF}1} -\mu^{\text{HF}} N_I)}}{\sum_J e^{-\beta (E_J^{\text{HF}1} - \mu^{\text{HF}} N_J)}}  \\
&=& \prod_i^{\text{occ.}} f_i^- \prod_a^{\text{vir.}} f_a^+ ,
\end{eqnarray}
where $f_p^\mp$ are given by Eqs.\ (\ref{FD:minus}) and (\ref{FD:plus}) with variational parameter $\epsilon_p^{\text{HF}}$ in the place of $\epsilon_p^{(0)}$
and $\mu^{\text{HF}}$ in the place of $\mu^{(0)}$. 

We then introduce a {\it second} Hamiltonian $\hat{H}$, which is exact.
\begin{eqnarray}
\hat{H} = \sum_p h_{pq} \hat{p}^\dagger \hat{q} + \frac{1}{4} \sum_{p,q,r,s} \langle pq || rs \rangle \hat{p}^\dagger\hat{q}^\dagger \hat{s} \hat{r}, \label{exactHamiltonian}
\end{eqnarray} 
where $h_{pq}$ is the one-electron (``core'') part of the Hamiltonian and $\langle pq||rs \rangle$ is an anti-symmetrized two-electron integral.\cite{szabo,shavitt}
It then gives a second definition of state energies as
\begin{eqnarray}
E_I^{\text{HF}2} = \langle I | \hat{H} | I \rangle = \sum_i^{\text{occ.}} h_{ii} 
+ \frac{1}{2} \sum_{i,j}^{\text{occ.}} \langle ij || ij \rangle , \label{HF:energy}
\end{eqnarray} 
where, as mentioned before, $i$ and $j$ run over spinorbitals occupied in the $I$th Slater determinant. 

We postulate $\Omega^{\text{HF}}$ by its density-matrix definition [Eq.\ (\ref{Exact:Omega2})] with a hybrid use of $\rho_I^{\text{HF}1}$ and $E_I^{\text{HF}2}$, 
\begin{eqnarray}
\Omega^{\text{HF}} &\equiv& \sum_I \rho_I^{\text{HF}1} (E_I^{\text{HF}2} - \mu^{\text{HF}} N_I) + \frac{1}{\beta} \sum_I \rho_I^{\text{HF}1} \ln \rho_I^{\text{HF}1} \nonumber\\ \label{HF:Omega1}  \\
&=& \sum_p ( h_{pp} -\mu^{\text{HF}}) f_p^- + \frac{1}{2} \sum_{p,q} \langle pq || pq \rangle f_p^- f_q^-  \nonumber\\
&& + \frac{1}{\beta} \sum_p \left( f_p^- \ln f_p^- + f_p^+ \ln f_p^+  \right)  \label{HF:Omega2}  \\
&\neq&-\frac{1}{\beta} \ln \Xi^{\text{HF}} , 
\end{eqnarray}
where the inequality indicates that no consistent grand partition function seems identifiable for this theory. 
The last equality [Eq.\ (\ref{HF:Omega2})] can be proven with the aid of Boltzmann-sum identities I and III of Ref.\ \onlinecite{HirataJhaJCP2020}.

Likewise, the internal energy and entropy are defined with the density matrix as
\begin{eqnarray}
U^{\text{HF}} & \equiv& \sum_I \rho_I^{\text{HF}1}E_I^{\text{HF}2} \label{HF:U1}  \\
&=& \sum_p h_{pp}  f_p^- + \frac{1}{2} \sum_{p,q} \langle pq || pq \rangle f_p^- f_q^-  \label{HF:U2} \\
&\neq&-\frac{\partial}{\partial \beta} \ln \Xi^{\text{HF}} + \mu^\text{HF}\bar{N}, \\
S^{\text{HF}} &\equiv& -k_\text{B} \sum_I \rho_I^{\text{HF}1} \ln \rho_I^{\text{HF}1} \\
&=&  \frac{1}{\beta} \sum_p \left( f_p^- \ln f_p^- + f_p^+ \ln f_p^+  \right) , \label{HF:S}  
\end{eqnarray}
where the inequality again underscores the nonexistence of a consistent grand partition function for $U^{\text{HF}}$.

Thermal HF theory stipulates that $\Omega^{\text{HF}}$ be minimized by varying $\epsilon_p^{\text{HF}}$. 
This is the same as varying $f_p^-$. 
Therefore, we demand
\begin{eqnarray}
0 &=& \frac{\partial \Omega^{\text{HF}}}{\partial f_p^-} \\
&=& h_{pp} - \mu^{\text{HF}} + \sum_q \langle pq||pq \rangle f_q^- 
+\frac{1}{\beta} \ln \frac{f_p^-}{f_p^+} \\
&=& h_{pp} - \mu^{\text{HF}} + \sum_q \langle pq||pq \rangle f_q^- - ( \epsilon_p^{\text{HF}} - \mu^{\text{HF}}) , \label{HF:minimization}
\end{eqnarray}
which is achieved by setting
\begin{eqnarray}
\epsilon_{pq}^{\text{HF}} &=& h_{pq} + \sum_r \langle pr || qr \rangle f_r^-  =  \delta_{pq}\epsilon_p^\text{HF}, 
\label{HF:epsilon} 
\end{eqnarray}
where $\delta_{pq}$ is the Kronecker delta. 
This is a well-known formula for thermal HF orbital energies, whose physical meaning is unknown,\cite{Pain} but proposed in the following article.\cite{Hirata24}
Thermal HF theory thus involves a unitary rotation of orbitals that makes $\bm{\epsilon}^\text{HF}$ diagonal. 

To reiterate key points, there is no single, consistent grand partition function for this theory from which 
all of its thermodynamic functions are derived. This is because the theory is postulated with a hybrid use of the two different Hamiltonians, one for the density matrix and 
the other for the state energies, and such a derivation cannot be accommodated by a single, consistent grand partition function. 
This raises a legitimate concern\cite{Kirzhnits} that thermal HF theory might not obey all fundamental thermodynamic relations. 
This concern may be exarcebated by the fact that the theory's $f_p^-$ contains $\epsilon_p^{\text{HF}}$, which in turn depends on
$f_p^-$, and therefore its derivative with respect to $\beta$ or $\mu^{\text{HF}}$ cannot be written in a closed form; it is rather  
a solution to a system of linear equations such as
\begin{eqnarray}
\frac{\partial f_p^-}{\partial \mu^{\text{HF}}} &=& \left(\beta - \beta \frac{\partial \epsilon_p^{\text{HF}}}{\partial \mu^{\text{HF}}}  \right) f_p^-f_p^+ 
\nonumber\\
&=& \left( \beta - \beta \sum_r \langle pr || pr \rangle \frac{\partial f_r^-}{\partial \mu^{\text{HF}}}  \right) f_p^-f_p^+ ,
\end{eqnarray}
or
\begin{eqnarray}
\frac{\partial f_p^-}{\partial \beta} &=& -\left( \epsilon_p^{\text{HF}} + \beta \frac{\partial \epsilon_p^{\text{HF}}}{\partial \beta} -\mu^{\text{HF}} \right) f_p^-f_p^+ 
\nonumber\\
&=& -\left( \epsilon_p^{\text{HF}} + \beta \sum_r \langle pr || pr \rangle \frac{\partial f_r^-}{\partial \beta} -\mu^{\text{HF}} \right) f_p^-f_p^+ .
\end{eqnarray}

However, as proven by Argyres {\it et al.},\cite{Argyres1974} thermal HF theory does obey all thermodynamic relations (insofar as they do not involve a grand partition function). 
Here, we provide an alternative, more direct proof of this claim by explicitly evaluating the relations.

Equation (\ref{Exact:OmegaUmuS}) is trivially satisfied.

That Eq.\ (\ref{Exact:relation1}) is also obeyed can be shown without explicit knowledge of the derivative of $f_p^-$ [as implied by Eq.\ (\ref{FD:relation1explicit})].
 \begin{eqnarray}
-\frac{\partial \Omega^{\text{HF}}}{\partial \mu^{\text{HF}}} &=& \sum_p f_p^- - \sum_{p,q} \langle pq || pq  \rangle \frac{\partial f_p^-}{\partial \mu^{\text{HF}}} f_q^-    
\nonumber\\&& - \sum_p (h_{pp} - \mu^{\text{HF}}) \frac{\partial f_p^-}{\partial \mu^{\text{HF}}} 
\nonumber\\&& 
 -\frac{1}{\beta} \sum_p \frac{\partial f_p^-}{\partial \mu^{\text{HF}}} \ln f_p^- -\frac{1}{\beta} \sum_p \frac{\partial f_p^-}{\partial \mu^{\text{HF}}}
 \nonumber\\&&
-\frac{1}{\beta} \sum_p \frac{\partial f_p^+}{\partial \mu^{\text{HF}}} \ln f_p^+ -\frac{1}{\beta} \sum_p \frac{\partial f_p^+}{\partial \mu^{\text{HF}}} \nonumber \\
\nonumber\\&=&
\sum_p f_p^-  - \sum_p \left( \epsilon_p^{\text{HF}} - \mu^{\text{HF}} \right) \frac{\partial f_p^-}{\partial \mu^{\text{HF}}}
\nonumber\\&&
 -\frac{1}{\beta} \sum_p \frac{\partial f_p^-}{\partial \mu^{\text{HF}}}  \ln \frac{f_p^-}{f_p^+}
 \nonumber\\&=& \sum_p f_p^-  = \bar{N},
\end{eqnarray}
where an equivalent of Eq.\ (\ref{lnfoverf}) was used. 
The last equality is the condition that determines $\mu^{\text{HF}}$, which is an integral part of the ansatz of thermal HF theory.

Likewise, Eq.\ (\ref{Exact:relation2}) is satisfied by the thermal HF quantities for we can show 
\begin{eqnarray}
-\frac{\partial \Omega^{\text{HF}}}{\partial T} &=& k_{\text{B}}\beta^2 \frac{\partial\Omega^{\text{HF}}}{\partial \beta} 
\nonumber\\
&=& k_{\text{B}}\beta^2 \sum_p (h_{pp} - \mu) \frac{\partial f_p^-}{\partial \beta}  
+ k_{\text{B}}\beta^2 \sum_{p,q} \langle pq || pq \rangle \frac{\partial f_p^-}{\partial \beta}  f_q^-
\nonumber\\
&& - k_{\text{B}}  \left( f_p^- \ln f_p^- + f_p^+ \ln f_p^+  \right)
\nonumber\\
&& + k_{\text{B}}\beta \sum_p \left(\frac{\partial f_p^-}{\partial \beta} \ln f_p^- + \frac{\partial f_p^-}{\partial \beta} + \frac{\partial f_p^+}{\partial \beta} \ln f_p^+ + \frac{\partial f_p^+}{\partial \beta} \right) 
\nonumber\\
&=& k_{\text{B}}\beta^2 \sum_p \left(\epsilon_p^{\text{HF}} - \mu^{\text{HF}}\right) \frac{\partial f_p^-}{\partial \beta} 
\nonumber\\&&
- k_{\text{B}}  \left( f_p^- \ln f_p^- + f_p^+ \ln f_p^+  \right)
\nonumber\\&&
 + k_{\text{B}}\beta \sum_p \frac{\partial f_p^-}{\partial \beta} \ln \frac{f_p^-}{f_p^+} 
\nonumber\\
&=& - k_{\text{B}}  \left( f_p^- \ln f_p^- + f_p^+ \ln f_p^+  \right) = S^{\text{HF}},
\end{eqnarray}
again without an explicit evaluation of $\partial f_p^- / \partial \beta$, as implied by Eq.\ (\ref{FD:relation2explicit}). 

The zero-temperature limit of $U^\text{HF}$ is the zero-temperature HF energy for the ground state. 
The $\epsilon_p^\text{HF}$ also reduces to the corresponding zero-temperature HF orbital energy as $T \to 0$.

\section{Thermal single-determinant approximation}

\subsection{TSDA0\label{sec:TSDA0}}

The thermal single-determinant approximation (TSDA) of Kaplan and Argyres\cite{Kaplan1975,Farid2000} 
is defined by the grand partition function that sums over all Slater-determinant states whose energies 
are evaluated with the exact Hamiltonian of Eq.\ (\ref{exactHamiltonian}). This is followed by minimization of the grand potential with respect to orbital rotation. 

Its grand partition function reads
\begin{eqnarray}
 \Xi^{\text{TSDA0}}  &\equiv& \sum_{I} e^{-\beta(E_I^{\text{TSDA0}} - \mu^{\text{TSDA0}} N_I)}, \label{TSDA0:Xi} 
\end{eqnarray}
where $E_I^{\text{TSDA0}}$ is defined with the exact Hamiltonian [Eq.\ (\ref{HF:energy})], i.e.,
\begin{eqnarray}
E_I^{\text{TSDA0}} 
&=& \langle I |\hat{H} | I\rangle = \sum_i^{\text{occ.}} \epsilon_i^{\text{TSDA0}} - \frac{1}{2} \sum_{i,j}^{\text{occ.}} \langle ij || ij \rangle \label{TSDA0:epsilon}
\end{eqnarray} 
with $\epsilon_p^{\text{TSDA0}}$ being the zero-temperature HF orbital energy,
\begin{eqnarray}
 \epsilon_p^{\text{TSDA0}} = h_{pp} + \sum_{j}^{\text{occ.}} \langle pj || pj \rangle ,
\end{eqnarray} 
whose physical meaning is provided by Koopmans' theorem.\cite{szabo} It 
should be distinguished from the thermal HF orbital energy of Eq.\ (\ref{HF:epsilon}); the $\epsilon_p^{\text{TSDA0}}$ is constant of $\mu^{\text{TSDA0}}$ or of $\beta$. 

Unlike Fermi--Dirac theory, this grand partition function does not simplify any further, as speculated by its inventors.\cite{Kaplan1975}
It is not the grand partition function of thermal HF theory, either. 
We can still evaluate this $\Xi^{\text{TSDA0}}$ by brute force, using a thermal FCI program,\cite{Kou} and also determine
$\Omega^{\text{TSDA0}}$, $U^{\text{TSDA0}}$, $\mu^{\text{TSDA0}}$, and $S^{\text{TSDA0}}$ with the following formulas:
\begin{eqnarray}
\Omega^{\text{TSDA0}} &\equiv&-\frac{1}{\beta} \ln \Xi^{\text{TSDA0}} , \\
U^{\text{TSDA0}} &\equiv& \sum_{I} \rho_I^{\text{TSDA0}}E_I^{\text{TSDA0}}, \\
\bar{N} &\equiv& \sum_I  \rho_I^{\text{TSDA0}}N_I, \label{TSDA0:mu}\\
S^{\text{TSDA0}} &\equiv& \frac{U^{\text{TSDA0}} - \Omega^{\text{TSDA0}} - \mu^{\text{TSDA0}} \bar{N}}{T}
\end{eqnarray}
with
\begin{eqnarray}
\rho_I^{\text{TSDA0}} &=& \frac{ e^{-\beta(E_I^{\text{TSDA0}} - \mu^{\text{TSDA0}} N_I)}}{\sum_{J} e^{-\beta(E_J^{\text{TSDA0}} - \mu^{\text{TSDA0}} N_J)}}.
\end{eqnarray}
Equation (\ref{TSDA0:mu}) determines $\mu^{\text{TSDA0}}$ based on $\bar{N}$. 
In our ansatz, we forgo the requirement that the grand potential be minimized. We call this method TSDA0, which differs from the original ansatz of Kaplan and Argyres\cite{Kaplan1975} 
in the absence of minimization. There are neither thermal orbital energies nor Fermi--Dirac-like distribution functions in this multi-determinant theory (but we still call it a ``mean-field'' theory).

While the practical utility of this method is severely limited, we have implemented TSDA0 and examined its numerical differences from the other thermal mean-field theories
considered in this study (see Sec.\ \ref{sec:numerical}). 

The thermodynamic relations of Eqs.\ (\ref{Exact:OmegaUmuS})--(\ref{Exact:relation2}) are obeyed by this theory by construction.
The zero-temperature limit of $U^\text{TSDA0}$ is the zero-temperature HF energy for the ground state. Owing to the trace invariance 
of the Hamiltonian matrix in the basis of all Slater determinants, TSDA0 converges at thermal FCI theory in the high-$T$ limit. 

\subsection{TSDA1\label{sec:TSDA1}} 

To render TSDA0 more practical, we make the following approximations:
Expanding the exponential of the two-electron part of the energy in a Taylor series, we have
\begin{widetext}
\begin{eqnarray}
e^{-\beta( E_I - \mu N_I)} &=& e^{-\beta\sum_{i} (\epsilon_i - \mu) } e^{\beta\sum_{i<j} \langle ij || ij \rangle} \nonumber \\
&=& e^{-\beta\sum_{i} (\epsilon_i - \mu) } 
\left\{ 1 + \sum_{i<j}b_{ij} + \frac{1}{2!} \left( \sum_{i<j}b_{ij} \right)^2 + \frac{1}{3!} \left( \sum_{i<j}b_{ij} \right)^3 + \dots  \right\}  \label{TSDA1:approx} 
\end{eqnarray}
with $b_{ij} = \beta \langle ij || ij \rangle$. 
The grand partition function of Eq.\ (\ref{TSDA0:Xi}) can then be reduced to a sum-over-orbitals expression,
\begin{eqnarray}
 \Xi^\text{TSDA1}  &\equiv& \sum_{I_0} 1 +\sum_{I_1} \sum_{i}^{I_1} e^{-\beta (\epsilon_{i}-\mu) } 
  + \sum_{I_2} \sum_{i < j}^{I_2} \left\{1 +  b_{ij} + \frac{1}{2!} (b_{ij})^2 + \frac{1}{3!} (b_{ij})^3+ \dots \right\} e^{-\beta (\epsilon_{i}-\mu)}e^{-\beta ( \epsilon_{j}-\mu) }   \nonumber\\
&& + \sum_{I_3} \sum_{i < j<k}^{I_3} \left\{1 + b_{ij} + b_{ik} + b_{jk} + \frac{1}{2!} (b_{ij} + b_{ik} + b_{jk})^2 + \frac{1}{3!} (b_{ij} + b_{ik} + b_{jk})^3 + \dots \right\} 
e^{-\beta (\epsilon_{i}-\mu) }e^{-\beta (\epsilon_{j}-\mu) }e^{-\beta (\epsilon_{k}-\mu) }  
+ \dots \nonumber\\
&\approx& \left\{ 1 +  \sum_{p < q} b_{pq} f_p^- f_q^- + \frac{1}{2!} \left( \sum_{p < q} b_{pq} f_p^- f_q^- \right)^2 + \frac{1}{3!} \left( \sum_{p < q} b_{pq} f_p^- f_q^- \right)^3 + \dots \right\} 
\prod_p \left( 1 + e^{-\beta (\epsilon_p-\mu) }\right)  \nonumber\\
&=& e^{\frac{1}{2} \sum_{p,q} b_{pq} f_p^- f_q^-}  \prod_p \left( 1 + e^{-\beta (\epsilon_p-\mu) }\right),
\end{eqnarray}
\end{widetext}
where $I_n$ denotes the set of all $n$-electron Slater determinants. In the approximate equality, Boltzmann-sum identify VI [Eq.\ (A6)] of Ref.\ \onlinecite{HirataJhaJCP2020} and its higher-order analogues 
were used. It is  approximate because only the sums with the ``no coinc.''\ index restrictions\cite{HirataJhaJCP2020} are retained because they are expected to be dominant, whereupon the restrictions are lifted. 
This is a lowest-order Ursell--Mayer cumulant expansion.\cite{Ursell1927,Mayer1941,Morita3,Andersen1977,Reichl}

The corresponding grand potential is then given by
\begin{eqnarray}
\Omega^{\text{TSDA1}} &\equiv& -\frac{1}{\beta} \ln \Xi^\text{TSDA1} \\
&=& \sum_p \left(\epsilon_p^{\text{TSDA1}}-\mu^{\text{TSDA1}}\right)  f_p^- - \frac{1}{2} \sum_{p, q} {\langle pq || pq \rangle} f_p^- f_q^- \nonumber \\
&& + \frac{1}{\beta} \sum_p \left( f_p^- \ln f_p^-+ f_p^+ \ln f_p^+\right) , \label{TSDA1:Omega} 
\end{eqnarray}
where $\epsilon_p^{\text{TSDA1}}$ is the $p$th spinorbital energy of a reference wave function. 
The $f_p^\mp$ are defined by Eqs.\ (\ref{FD:minus}) and (\ref{FD:plus}) with $\epsilon_p^{\text{TSDA1}}$ in the place 
of $\epsilon_p^{(0)}$ and $\mu^{\text{TSDA1}}$ in the place of $\mu^{(0)}$. 
The $\mu^{\text{TSDA1}}$ is determined by 
\begin{eqnarray}
\bar{N} &\equiv& \frac{1}{\beta}\frac{\partial}{\partial \mu^\text{TSDA1}} \ln \Xi^\text{TSDA1} = \sum_p f_p^-. \label{TSDA1:mu}
\end{eqnarray}
The internal energy is defined by
\begin{eqnarray}
U^{\text{TSDA1}} &\equiv& -\frac{\partial}{\partial \beta} \ln \Xi^\text{TSDA1} +\mu^{\text{TSDA1}} \bar{N}  \\
&=&  \sum_p \epsilon_p^{\text{TSDA1}} f_p^- - \frac{1}{2} \sum_{p, q} {\langle pq || pq \rangle} f_p^- f_q^-.
\end{eqnarray}
We call this method TSDA1. 

These formulas are the same as the counterparts of thermal HF theory. If one stipulates, in the spirit of the original ansatz of Kaplan and Argyres,\cite{Kaplan1975} 
that  $\Omega^{\text{TSDA1}}$ [Eq.\ (\ref{TSDA1:Omega})] be minimized by varying $\epsilon_p^{\text{TSDA1}}$, 
TSDA1 exactly reverts to thermal HF theory and does not constitute a new ansatz. 
Instead, starting with the TSDA ansatz of Kaplan and Argyres and invoking the cumulant expansion can be viewed as a second way of deriving or rationalizing 
the thermodynamic functions of thermal HF theory from a grand partition function. In Sec.\ \ref{sec:PT}, we offer a third derivation based on another grand partition function.  

If one employed a partitioning alternative to Eq.\ (\ref{TSDA1:approx}),
\begin{eqnarray}
e^{-\beta( E_I - \mu N_I)} &=& e^{-\beta\sum_{i} (h_{ii} - \mu) } e^{-\beta\sum_{i<j} \langle ij || ij \rangle} ,
\end{eqnarray}
the correspondence with thermal HF theory would be lost. We shall not consider this ansatz.

The TSDA1 does not obey thermodynamic relations unless $\Omega^{\text{TSDA1}}$ is minimized to become equal to $\Omega^{\text{HF}}$. 
For example,
 \begin{eqnarray}
-\frac{\partial \Omega^{\text{TSDA1}}}{\partial \mu^{\text{TSDA1}}} 
&=&  \sum_p f_p^- + \sum_{p,q} \langle pq || pq  \rangle \frac{\partial f_p^-}{\partial \mu^{\text{TSDA1}}} f_q^-    
\nonumber\\&&
- \sum_p \left (\epsilon_p^{\text{TSDA1}} - \mu^{\text{TSDA1}} \right ) \frac{\partial f_p^-}{\partial \mu^{\text{TSDA1}}} 
\nonumber\\&& 
 -\frac{1}{\beta} \sum_p \frac{\partial f_p^-}{\partial \mu^{\text{TSDA1}}} \ln f_p^- -\frac{1}{\beta} \sum_p \frac{\partial f_p^-}{\partial \mu^{\text{TSDA1}}}
 \nonumber\\&&
-\frac{1}{\beta} \sum_p \frac{\partial f_p^+}{\partial \mu^{\text{TSDA1}}} \ln f_p^+ -\frac{1}{\beta} \sum_p \frac{\partial f_p^+}{\partial \mu^{\text{TSDA1}}} \nonumber \\
\nonumber\\&=&
\sum_p f_p^- + \sum_{p,q} \langle pq || pq  \rangle \frac{\partial f_p^-}{\partial \mu^{\text{TSDA1}}} f_q^-
 \nonumber\\&&
  - \sum_p \left( \epsilon_p^{\text{TSDA1}} - \mu^{\text{TSDA1}} \right) \frac{\partial f_p^-}{\partial \mu^{\text{TSDA1}}}
\nonumber\\&&
 -\frac{1}{\beta} \sum_p \frac{\partial f_p^-}{\partial \mu^{\text{TSDA1}}}  \ln \frac{f_p^-}{f_p^+}
 \nonumber\\&=& \sum_p f_p^-  + \sum_{p,q} \langle pq || pq  \rangle \frac{\partial f_p^-}{\partial \mu^{\text{TSDA1}}} f_q^- \neq \bar{N}.
\end{eqnarray}
Nor does it satisfy Eq.\ (\ref{Exact:relation2}) unless $\Omega^{\text{TSDA1}}$ is minimized. 


\section{First-order finite-temperature many-body perturbation theory\label{sec:PT}}

Finite-temperature MBPT (Ref.\ \onlinecite{HirataJha,HirataJhaJCP2020,Hirata2021}) partitions the exact Hamiltonian of Eq.\ (\ref{exactHamiltonian}) into the zeroth-order part and perturbation,
\begin{eqnarray}
\hat{H} = \hat{H}_0 + \hat{V},
\end{eqnarray}
where $\hat{H}_0$ takes the form of Eq.\ (\ref{H0}). Hence, $\epsilon_p^{(0)}$ in it is the $p$th orbital energy of an arbitrary reference Slater determinant.
The corresponding zeroth-order thermodynamic functions are those of Fermi--Dirac theory, 
satisfying all thermodynamic relations. The $f_p^\mp$ are defined precisely by Eqs.\ (\ref{FD:minus}) and (\ref{FD:plus}). 

The corrections of first-order finite-temperature MBPT or finite-temperature MBPT(1) 
have been derived in Refs.\ \onlinecite{HirataJha,HirataJhaJCP2020,Hirata2021} and they read
\begin{eqnarray}
\Omega^{(1)} &=& \sum_p \left( \epsilon_p^{\text{HF}} - \epsilon_p^{(0)} -  \mu^{(1)} \right) f_p^- 
 -\frac{1}{2} \sum_{p,q} \langle pq || pq  \rangle  f_p^-f_q^-, \nonumber\\
 \label{Omega1} \\
U^{(1)} &=&  \sum_p \left( \epsilon_p^{\text{HF}} - \epsilon_p^{(0)} \right) f_p^-  -\frac{1}{2} \sum_{p,q} \langle pq || pq  \rangle  f_p^-f_q^- \nonumber\\
&& -\beta \sum_p \left( \epsilon_p^{\text{HF}} - \epsilon_p^{(0)} -  \mu^{(1)} \right) \epsilon_p^{(0)} f_p^- f_p^+,  \label{U1}\\
\mu^{(1)} &=& \frac{ \sum_p \left( \epsilon_p^{\text{HF}} - \epsilon_p^{(0)} \right) f_p^- f_p^+}{\sum_p f_p^- f_p^+} , \label{mu1} \\
S^{(1)} &=& \frac{U^{(1)} - \Omega^{(1)} - \mu^{(1)} \bar{N}} {T} \nonumber\\
&=&- {k_{\text{B}}\beta^2} \sum_p \left( \epsilon_p^{\text{HF}} - \epsilon_p^{(0)} -  \mu^{(1)} \right) \epsilon_p^{(0)} f_p^- f_p^+, \label{S1}
\end{eqnarray}
where $\epsilon_p^{\text{HF}}$ is given by Eq.\ (\ref{HF:epsilon}) evaluated for the zeroth-order orbitals; it has the same functional form as 
the thermal HF orbital energy, but generally has different numerical values.

As pointed out earlier,\cite{Hirata2021} thermal HF theory's $\Omega^{\text{HF}}$ [Eq.\ (\ref{HF:Omega2})] does not coincide with $\Omega^{(0)} + \Omega^{(1)}$ [Eqs.\ (\ref{FD:Omega})+(\ref{Omega1})]. Nor does
$U^{\text{HF}}$ [Eq.\ (\ref{HF:U2})] correspond to $U^{(0)} + U^{(1)}$ [Eqs.\ (\ref{FD:U})+(\ref{U1})]. This is unlike zero-temperature HF theory, whose energy expression is the first-order 
M{\o}ller--Plesset perturbation energy expression.\cite{szabo,shavitt}

The thermodynamic relation of Eq.\ (\ref{Exact:OmegaUmuS}) is trivially satisfied. 
The thermodynamic relation of Eq.\ (\ref{Exact:relation1}) now becomes
\begin{eqnarray}
\bar{N} = -\frac{\partial}{\partial \mu^{(0)}} \left( \Omega^{(0)} + \Omega^{(1)} \right),
\end{eqnarray}
which is broken down into order-by-order conditions,
\begin{eqnarray}
\bar{N} &=& -\frac{\partial\Omega^{(0)}}{\partial \mu^{(0)}} , \\
0 &=& -\frac{\partial\Omega^{(1)}}{\partial \mu^{(0)}} . \label{PT1:relation1}
\end{eqnarray}

The first of these is obeyed by Fermi--Dirac theory [Eq.\ (\ref{FD:relation1explicit})]. That the second  is also satisfied is confirmed as follows:
 \begin{eqnarray}
-\frac{\partial \Omega^{(1)}}{\partial \mu^{(0)}} &=& - \sum_p \left( \epsilon_p^{\text{HF}} - \epsilon_p^{(0)} -  \mu^{(1)} \right) \frac{\partial f_p^-}{\partial \mu^{(0)}}  
\nonumber\\&&
- \sum_{p,q} \langle pq || pq  \rangle \frac{\partial f_q^-}{\partial \mu^{(0)}} f_p^-  +  \sum_{p,q} \langle pq || pq  \rangle \frac{\partial f_p^-}{\partial \mu^{(0)}} f_q^- 
\nonumber\\&=&
- \sum_p \left( \epsilon_p^{\text{HF}} - \epsilon_p^{(0)} -  \mu^{(1)} \right)\beta f_p^-f_p^+ 
 \nonumber\\&=& 0, \label{MP1:relation1}
\end{eqnarray}
where Eq.\ (\ref{mu_deriv}) was used in the penultimate equality and the definition of $\mu^{(1)}$ [Eq.\ (\ref{mu1})] in the last equality.
Note that it would be misleading to write $\mu^{(1)}\bar{N}$ for  $\sum_p \mu^{(1)}f_p^-$ in the definition of $\Omega^{(1)}$ [Eq.\ (\ref{Omega1})] in this context. 
It would be also wrong to replace $\mu^{(1)}$ in $\Omega^{(1)}$ [Eq.\ (\ref{Omega1})] by its formula [Eq.\ (\ref{mu1})] because $\mu^{(1)}$ is held fixed when 
the partial derivative with respect to $\mu^{(0)}$ is taken. In this case, the explicit derivative of $f_p^\mp$ [Eq.\ (\ref{mu_deriv})] is needed to prove the above identity.

Owing to Eq.\ (\ref{FD:relation2explicit}), the thermodynamic relation of Eq.\ (\ref{Exact:relation2}) would be satisfied if the following was obeyed: 
\begin{eqnarray}
S^{(1)} &=& -\frac{\partial \Omega^{(1)}}{\partial T}. \label{PT1:relation2}
\end{eqnarray}
That this is indeed the case can also be confirmed explicitly.
\begin{eqnarray}
-\frac{\partial \Omega^{(1)}}{\partial T} &=& k_{\text{B}}\beta^2 \frac{\partial\Omega^{(1)}}{\partial \beta} 
\nonumber\\
&=& k_{\text{B}}\beta^2 \sum_p \left( \epsilon_p^{\text{HF}} - \epsilon_p^{(0)} -  \mu^{(1)} \right) \frac{\partial f_p^-}{\partial \beta} 
\nonumber\\ &&
+ k_{\text{B}}\beta^2 \sum_{p,q} \langle pq || pq \rangle \frac{\partial f_q^-}{\partial \beta}  f_p^-
\nonumber\\ &&
- k_{\text{B}}\beta^2 \sum_{p,q} \langle pq || pq \rangle \frac{\partial f_p^-}{\partial \beta}  f_q^-
\nonumber\\
&=& - k_{\text{B}}\beta^2 \sum_p \left( \epsilon_p^{\text{HF}} - \epsilon_p^{(0)} -  \mu^{(1)} \right) \left(\epsilon_p^{(0)} - \mu^{(0)}\right) f_p^-f_p^+  
\nonumber\\
&=& S^{(1)}, \label{MP1:relation2}
\end{eqnarray}
where Eq.\ (\ref{beta_deriv}) was used in the penultimate equality and Eq.\ (\ref{MP1:relation1}) in the last equality. 

Therefore, all fundamental thermodynamic relations are satisfied by finite-temperature MBPT(1). 
This is expected in view of its construction based on such relations.\cite{HirataJha,HirataJhaJCP2020,Hirata2021}
In fact, it was pointed out in 
Ref.\ \onlinecite{HirataJhaJCP2020} that an expedient derivation of $\mu^{(1)}$ and $U^{(1)}$ would be to start with
the identities,
\begin{eqnarray}
\left( \frac{\partial \Omega^{(1)}}{\partial \mu^{(0)}}\right)_{\mu^{(1)}} &=& 0, \\
U^{(1)} &=& \Omega^{(1)} + \mu^{(1)} \bar{N} + \beta \left( \frac{\partial \Omega^{(1)}}{\partial \beta}\right)_{\mu^{(0)},\mu^{(1)}} ,
\end{eqnarray}
which are equivalent to the thermodynamic relations of Eqs.\ (\ref{PT1:relation1}) and (\ref{PT1:relation2}), respectively, which must therefore be trivially satisfied.

When a thermal HF solution is used as the reference,
\begin{eqnarray}
\epsilon_p^{(0)} &\equiv& \epsilon_p^{\text{HF}} . \label{HFequalMP1}
\end{eqnarray}
This immediately implies
\begin{eqnarray}
\mu^{(1)} &=& 0, \label{MP1:THFequiv1}\\
S^{(1)} &=& 0,\label{MP1:THFequiv2}
\end{eqnarray}
and
\begin{eqnarray}
\Omega^{(1)} = U^{(1)} =  -\frac{1}{2} \sum_{p,q} \langle pq || pq  \rangle  f_p^-f_q^-. \label{MP1:THFequiv3}
\end{eqnarray}
Only with the thermal HF reference do $\Omega^{\text{HF}}$ and $U^{\text{HF}}$ coincide with $\Omega^{(0)} + \Omega^{(1)}$ and $U^{(0)} + U^{(1)}$, respectively,
both formalism-wise and numerically. 

Therefore, a third way of deriving the formulas (if not the ansatz) of thermal HF theory from a consistent grand partition function 
is to demand $\mu^{(1)} = 0$  and $S^{(1)} = 0$ in finite-temperature MBPT (which is rigorously derived from a perturbation expansion of the exact grand partition function\cite{HirataJha,HirataJhaJCP2020,Hirata2021}). 
These two conditions can be rewritten as orthogonality conditions among three vectors,
\begin{eqnarray}
\bm{a}\cdot \bm{b} &=& 0, \\
\bm{a}\cdot \bm{c} &=& 0,
\end{eqnarray}
with
\begin{eqnarray}
a_p &=& \epsilon_p^{\text{HF}} - \epsilon_p^{(0)}, \\
b_p &=& f_p^- f_p^+, \\
c_p &=& \epsilon_p^{(0)} f_p^- f_p^+.
\end{eqnarray}
Since $\bm{b} \nparallel \bm{c}$, we have $\bm{a} = \bm{0}$, which then implies Eq.\ (\ref{HFequalMP1}) and thus leads to the expressions of thermal HF theory. 

Furthermore, demanding $\mu^{(1)} = 0$ [Eq.\ (\ref{MP1:relation1})] and $S^{(1)} = 0$ [Eq.\ (\ref{MP1:relation2})] is equivalent to minimizing $\Omega^{\text{HF}}$. This can be understood by noting 
\begin{eqnarray}
0 &=& \frac{\partial\Omega^{\text{HF}} }{\partial f_p^-} =  \frac{\partial}{\partial f_p^-} \left( \Omega^{(0)} + \Omega^{(1)} \right)  =  \frac{\partial \Omega^{(1)} }{\partial f_p^-}\nonumber\\
&=& \frac{\partial \mu^{(0)} }{\partial f_p^-}\frac{\partial \Omega^{(1)} }{\partial \mu^{(0)} } + \frac{\partial T }{\partial f_p^-}\frac{\partial \Omega^{(1)} }{\partial T }  ,
\end{eqnarray}
where Eq.\ (\ref{FD:variational}) was used in the penultimate equality and any variation in $f_p^-$ can be actuated by variations in $\mu^{(0)}$ and $T$. 

The sum  $U^{(0)}+U^{(1)}$  converges at the ground-state energy of the first-order Hirschfelder--Certain degenerate perturbation theory\cite{hirschfelder} as $T \to 0$ unless the Kohn--Luttinger nonconvergence problem arises.\cite{Hirata_KL2021,Hirata_KL2022} In the thermal HF reference, $U^{(0)}+U^{(1)}$ correctly reduces to the zero-temperature HF energy for the ground state even under the same condition
as per the Luttinger--Ward prescription.\cite{luttingerward,Hirata_KL2021,Hirata_KL2022}

\section{Numerical Comparisons\label{sec:numerical}}

\begin{table*}
\caption{\label{table:Omega} Grand potential $\Omega$ (in $E_\text{h}$) of an ideal gas of the identical hydrogen fluoride molecules (0.9168\,\AA) in the STO-3G basis set as a function of temperature ($T$).}
\begin{ruledtabular}
\begin{tabular}{crrrrrrrrr}
& \multicolumn{4}{c}{Zero-temperature HF reference} & \multicolumn{4}{c}{Thermal HF reference} & \\ \cline{2-5} \cline{6-9} 
$T$ / K & \multicolumn{1}{c}{FD\tablenotemark[1]} & \multicolumn{1}{c}{TSDA0\tablenotemark[2]} & \multicolumn{1}{c}{TSDA1\tablenotemark[3]} & \multicolumn{1}{c}{MBPT(1)\tablenotemark[4]} 
            & \multicolumn{1}{c}{FD\tablenotemark[1]} & \multicolumn{1}{c}{TSDA0\tablenotemark[2]}  & \multicolumn{1}{c}{TSDA1=HF\tablenotemark[5]}& \multicolumn{1}{c}{MBPT(1)\tablenotemark[4]} & \multicolumn{1}{c}{FCI\tablenotemark[6]} \\ \hline
$10^4$ & $-53.51172$   & $-99.50757$   &  $-99.50758$  &  $-99.50758$  &   $-53.51172$ &   $-99.50757$ &   $-99.50758$ &   $-99.50758$ &   $-99.94377$ \\
$10^5$ & $-55.63656$   & $-101.66865$  & $-101.01485$  & $-100.90498$  &   $-55.33414$ &  $-101.92236$ &  $-101.02137$ &  $-101.02137$ &  $-102.10659$ \\
$10^6$ & $-105.94753$  & $-151.09870$  & $-150.09718$  & $-150.47317$  &  $-106.34446$ &  $-151.20266$ &  $-150.56294$ &  $-150.56294$ &  $-151.24440$ \\
$10^7$ & $-686.70814$  & $-730.06988$  & $-729.51682$  & $-729.90725$  &  $-687.10484$ &  $-730.08734$ &  $-729.93806$ &  $-729.93806$ &  $-730.09519$ \\
$10^8$ & $-6804.99036$ & $-6846.99928$ & $-6846.91697$ & $-6846.97502$ & $-6805.04985$ & $-6847.00152$ & $-6846.98049$ & $-6846.98049$ & $-6847.00247$ \\
\end{tabular}
\tablenotetext[1]{Fermi--Dirac theory with respective references.}
\tablenotetext[2]{TSDA0 (Sec.\ \ref{sec:TSDA0}) with respective references.}
\tablenotetext[3]{TSDA1 (Sec.\ \ref{sec:TSDA1}) with $\epsilon_p^{\text{TSDA1}}$ being the zero-temperature HF orbital energy.}
\tablenotetext[4]{Finite-temperature MBPT(1)\cite{HirataJha,HirataJhaJCP2020,Hirata2021} with respective references.} 
\tablenotetext[5]{TSDA1 with thermal HF reference or variational TSDA1 (Sec.\ \ref{sec:TSDA1}) is equivalent to thermal HF theory.}
\tablenotetext[6]{Thermal FCI theory,\cite{Kou} whose results are invariant with reference.}
\end{ruledtabular} 
\end{table*}

Table \ref{table:Omega} compares the grand potential $\Omega$ of an ideal gas of the identical hydrogen fluoride molecules (the bond length of 0.9168\,\AA) in the STO-3G basis set obtained by
various methods as a function of temperature ($T$). The methods considered are Fermi--Dirac theory (denoted by FD), the two modifications of the TSDA of Kaplan and Argyres\cite{Kaplan1975} 
(TSDA0 and TSDA1), finite-temperature MBPT(1)\cite{ HirataJha,HirataJhaJCP2020,Hirata2021} starting from either the zero-temperature HF reference (``zHF ref.''\ in 
figures) or  thermal HF reference (``tHF ref.'')\ using one and the same $T$. The thermal FCI results\cite{Kou} are also listed as the exact (within a basis set) benchmarks that are invariant with the reference.

The TSDA1 with the thermal HF reference (or the variational TSDA1)  is synonymous with thermal HF theory, and therefore, 
the corresponding two columns are consolidated into one (``TSDA1=HF'') in each table.  
The sum of the zeroth- and first-order finite-temperature MBPT corrections to $\Omega$ with the thermal HF reference is also equal to thermal HF theory as per Eqs.\ (\ref{MP1:THFequiv1})--({\ref{MP1:THFequiv3}), but they are shown separately because they go through independent formalisms and algorithms, mutually verifying each other.

Fermi--Dirac theory is  erroneous at lower $T$ for it neglects two-electron interactions altogether. 
At higher $T$, errors become relatively small as the entropy contributions dominate over two-electron-interaction energies. 
However, starting from Fermi--Dirac theory as the reference, finite-temperature MBPT(1) 
nearly completely erases the large errors and brings $\Omega$ in line with the other methods. With the thermal HF reference, thermal HF theory, TSDA1, and finite-temperature MBPT(1) 
give the identical values of $\Omega$, numerically verifying the aforementioned analytical expectations.

Excepting Fermi--Dirac theory and thermal FCI theory, all methods approach the same $T=0$ limit of $\Omega$, which is the ground-state HF energy minus 
$\bar{N}$ times the $T=0$ limit of $\mu$. The latter, in turn, corresponds to the midpoint\cite{Hirata_KL2021} of the highest-occupied (HOMO) and lowest-unoccupied molecular-orbital (LUMO) energies
of zero-temperature HF theory. At $T=0$, entropy is zero in all methods, obeying the third law of thermodynamics. 
The correct (FCI) $T=0$ limit of $\Omega$ is lower because the exact ground-state energy is more negative and 
the exact $\mu$ at $T=0$ is more positive. 

\begin{figure}
  \includegraphics[width=\columnwidth]{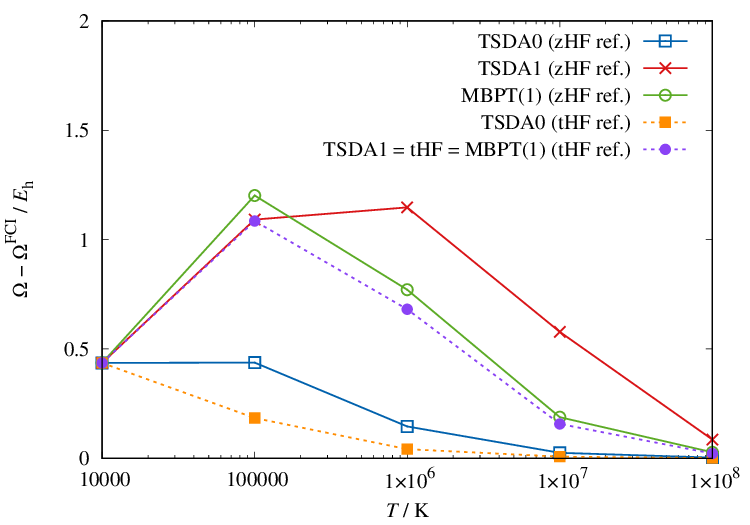}
\caption{The deviation from thermal FCI theory in grand potential $\Omega$ of an ideal gas of the identical hydrogen fluoride molecules (0.9168\,\AA) in the STO-3G basis set as a function of temperature ($T$). The ``zHF ref.''\ stands 
for the zero-temperature HF reference and ``tHF ref.''\ denotes the thermal HF reference. }
\label{fig:FH_Omega}
\end{figure}

Figure \ref{fig:FH_Omega} plots the deviation in $\Omega$ from the FCI value as a function of $T$. Fermi--Dirac theory is not shown as it is outside the graph. 
The following observations can be made: For each method, the thermal HF reference brings slightly better performance than the zero-temperature HF reference. 
The slight advantage is lost at both lower and higher temperatures. This is understandable given the same zero- and high-$T$ limits of most of these methods.

Finite-temperature MBPT(1) is stable with respect to the reference choice. The TSDA0 for $\Omega$ is slightly more dependent on the reference, but is much more accurate 
than thermal HF theory or finite-temperature MBPT(1). This is because the TSDA0 differs from thermal FCI theory only in its neglect of off-diagonal elements of the FCI Hamiltonian matrix; despite
its ``mean-field'' designation in this article, the TSDA0 is expected to take account of a substantial portion of electron-correlation effects. 
Whether its good performance persists in a bigger molecule or a more realistic basis set remains unknown, but this may be a moot point because the method is too expensive.

The TSDA1 can be viewed as a non-converged thermal HF theory and is much less stable with the reference choice than finite-temperature MBPT(1), underscoring 
the importance of $\mu^{(1)}$ and $S^{(1)}$, which are neglected in the TSDA1. Or put another way, minimizing $\Omega^\text{HF}$ in thermal HF theory is the same as requiring
$\mu^{(1)} = 0$ and $S^{(1)} = 0$ (see Sec.\ \ref{sec:PT}), eliminating the reference dependency.

\begin{table*}
\caption{\label{table:U} Same as Table \ref{table:Omega} but for internal energy $U$ (in $E_\text{h}$). The ground-state HF and FCI energies are  $-98.57076\,E_\text{h}$ and $-98.59659\,E_\text{h}$, respectively.}
\begin{ruledtabular}
\begin{tabular}{crrrrrrrrr}
& \multicolumn{4}{c}{Zero-temperature HF reference} & \multicolumn{4}{c}{Thermal HF reference} & \\ \cline{2-5} \cline{6-9} 
$T$ / K & \multicolumn{1}{c}{FD\tablenotemark[1]} & \multicolumn{1}{c}{TSDA0\tablenotemark[2]} & \multicolumn{1}{c}{TSDA1\tablenotemark[3]} & \multicolumn{1}{c}{MBPT(1)\tablenotemark[4]} 
            & \multicolumn{1}{c}{FD\tablenotemark[1]} & \multicolumn{1}{c}{TSDA0\tablenotemark[2]}  & \multicolumn{1}{c}{TSDA1=HF\tablenotemark[5]}& \multicolumn{1}{c}{MBPT(1)\tablenotemark[4]} & \multicolumn{1}{c}{FCI\tablenotemark[6]} \\ \hline
$10^4$ & $-52.57490$ & $-98.57076$ & $-98.57076$ & $-98.57076$ & $-52.57490$ & $-98.57076$ & $-98.57076$ & $-98.57076$ & $-98.59658$ \\
$10^5$ & $-52.01659$ & $-98.01569$ & $-97.39489$ & $-97.96445$ & $-52.25662$ & $-98.02133$ & $-97.94385$ & $-97.94385$ & $-98.04938$ \\
$10^6$ & $-50.59635$ & $-96.92316$ & $-94.74600$ & $-96.77300$ & $-52.57562$ & $-96.93579$ & $-96.79410$ & $-96.79410$ & $-96.94534$ \\
$10^7$ & $-45.78911$ & $-92.05181$ & $-88.59779$ & $-92.02465$ & $-49.19450$ & $-92.05407$ & $-92.02773$ & $-92.02773$ & $-92.05557$ \\
$10^8$ & $-42.36405$ & $-88.48687$ & $-84.29066$ & $-88.48208$ & $-46.55202$ & $-88.48720$ & $-88.48266$ & $-88.48266$ & $-88.48740$ \\
\end{tabular}
\tablenotetext[1]{$^{\text{-- f}}$ See the corresponding footnotes of Table \ref{table:Omega}.}
\end{ruledtabular} 
\end{table*}

\begin{figure}
  \includegraphics[width=\columnwidth]{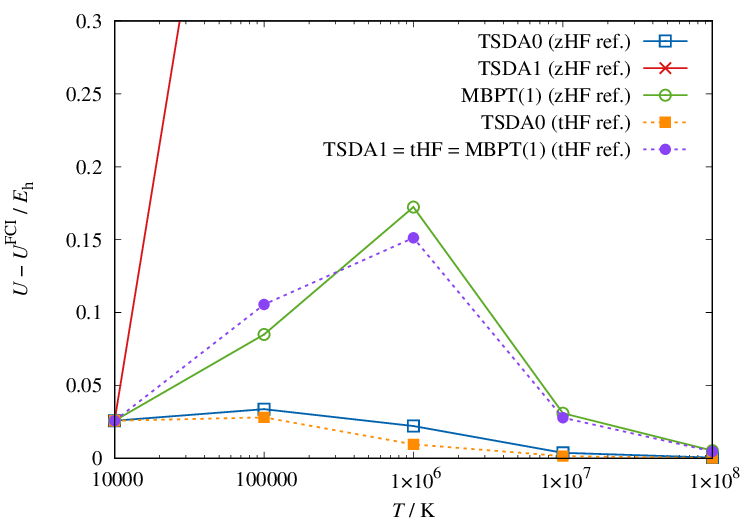}
\caption{Same as Fig.\ \ref{fig:FH_Omega} but for internal energy $U$.}
\label{fig:FH_U}
\end{figure}

Table \ref{table:U} compiles the internal energy $U$ of the same system as a function of $T$. Figure \ref{fig:FH_U} gives a graphical representation of the data 
as the deviations from thermal FCI benchmark. 
They numerically confirm that the TSDA1 and finite-temperature MBPT(1)
with the thermal HF reference are identified with thermal HF theory. The TSDA0, TSDA1, thermal HF theory, and finite-temperature MBPT(1), regardless of the reference, 
all have the same $T=0$ limits of the ground-state HF energy. The $T=0$ limit of the thermal FCI $U$ is the ground-state FCI energy and is certainly lower. 
Excepting Fermi--Dirac theory and the TSDA1 with the zero-temperature HF reference, the high-$T$ limits are also the same. The TSDA1
has a vastly different high-$T$ limit, which seems to be coupled with its poor performance for $\mu$ at high $T$ (see below). 
The TSDA0 and finite-temperature MBPT(1) are relatively insensitive to the reference, and the former is more accurate than the latter across a range of $T$.

\begin{table*}
\caption{\label{table:mu} Same as Table \ref{table:Omega} but for chemical potential $\mu$ (in $E_\text{h}$).}
\begin{ruledtabular}
\begin{tabular}{crrrrrrrrr}
& \multicolumn{4}{c}{Zero-temperature HF reference} & \multicolumn{4}{c}{Thermal HF reference} & \\ \cline{2-5} \cline{6-9} 
$T$ / K & \multicolumn{1}{c}{FD\tablenotemark[1]} & \multicolumn{1}{c}{TSDA0\tablenotemark[2]} & \multicolumn{1}{c}{TSDA1\tablenotemark[3]} & \multicolumn{1}{c}{MBPT(1)\tablenotemark[4]} 
            & \multicolumn{1}{c}{FD\tablenotemark[1]} & \multicolumn{1}{c}{TSDA0\tablenotemark[2]}  & \multicolumn{1}{c}{TSDA1=HF\tablenotemark[5]}& \multicolumn{1}{c}{MBPT(1)\tablenotemark[4]} & \multicolumn{1}{c}{FCI\tablenotemark[6]} \\ \hline
$10^4$ &   $0.09368$ &   $0.09368$ &   $0.09368$ &   $0.09368$ &   $0.09368$ &   $0.09368$ &   $0.09368$ &   $0.09368$ &   $0.13472$ \\
$10^5$ &   $0.27224$ &   $0.25534$ &   $0.27224$ &   $0.19705$ &   $0.20722$ &   $0.27949$ &   $0.20722$ &   $0.20722$ &   $0.29568$ \\
$10^6$ &   $3.96130$ &   $3.84656$ &   $3.96130$ &   $3.79234$ &   $3.80022$ &   $3.85625$ &   $3.80022$ &   $3.80022$ &   $3.85990$ \\
$10^7$ &  $47.15012$ &  $46.86660$ &  $47.15012$ &  $46.85201$ &  $46.85490$ &  $46.86822$ &  $46.85490$ &  $46.85490$ &  $46.86892$ \\
$10^8$ & $505.06450$ & $504.65447$ & $505.06450$ & $504.65229$ & $504.65280$ & $504.65468$ & $504.65280$ & $504.65280$ & $504.65476$ \\
\end{tabular}
\tablenotetext[1]{$^{\text{-- f}}$ See the corresponding footnotes of Table \ref{table:Omega}.}
\end{ruledtabular} 
\end{table*}

\begin{figure}
  \includegraphics[width=\columnwidth]{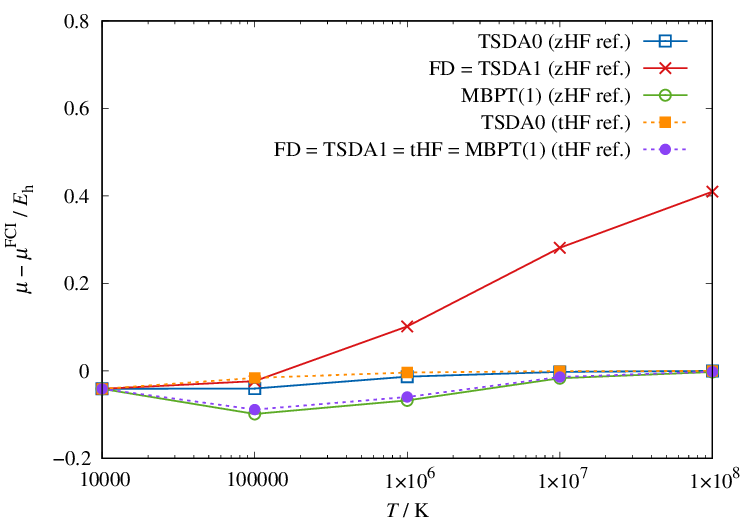}
\caption{Same as Fig.\ \ref{fig:FH_Omega} but for chemical potential $\mu$.}
\label{fig:FH_mu}
\end{figure}

Table \ref{table:mu} and Fig.\ \ref{fig:FH_mu} show the chemical potential $\mu$ as a function of $T$. For this thermodynamic function, Fermi--Dirac theory is reasonable; with 
the zero-temperature HF reference, it is the same as the TSDA1; with the thermal HF reference, it is equivalent to  the TSDA1, thermal HF theory, and 
finite-temperature MBPT(1). All mean-field theories considered (i.e., except for thermal FCI theory) approach the same $T=0$ limit, which is the average of the HOMO
and LUMO energies, but differs from the  $T=0$ limit of exact (thermal FCI) $\mu$. In the high-$T$ limit, all methods including thermal FCI theory approach the same limit except 
Fermi--Dirac theory and the TSDA1 in the zero-temperature HF reference. 

\begin{table*}
\caption{\label{table:S} Same as Table \ref{table:Omega} but for entropy $S$ (in $k_\text{B}$).}
\begin{ruledtabular}
\begin{tabular}{crrrrrrrrr}
& \multicolumn{4}{c}{Zero-temperature HF reference} & \multicolumn{4}{c}{Thermal HF reference} & \\ \cline{2-5} \cline{6-9} 
$T$ / K & \multicolumn{1}{c}{FD\tablenotemark[1]} & \multicolumn{1}{c}{TSDA0\tablenotemark[2]} & \multicolumn{1}{c}{TSDA1\tablenotemark[3]} & \multicolumn{1}{c}{MBPT(1)\tablenotemark[4]} 
            & \multicolumn{1}{c}{FD\tablenotemark[1]} & \multicolumn{1}{c}{TSDA0\tablenotemark[2]}  & \multicolumn{1}{c}{TSDA1=HF\tablenotemark[5]}& \multicolumn{1}{c}{MBPT(1)\tablenotemark[4]} & \multicolumn{1}{c}{FCI\tablenotemark[6]} \\ \hline
$10^4$ &   $0.00000$ &   $0.00008$ &   $0.00000$ &   $0.00000$ &   $0.00000$ &   $0.00008$ &   $0.00000$ &   $0.00000$ &   $0.00011$ \\
$10^5$ &   $2.83443$ &   $3.47228$ &   $2.83443$ &   $3.06324$ &   $3.17451$ &   $3.49286$ &   $3.17451$ &   $3.17451$ &   $3.47472$ \\
$10^6$ &   $4.96972$ &   $4.96079$ &   $4.96972$ &   $4.98189$ &   $4.97871$ &   $4.95905$ &   $4.97871$ &   $4.97871$ &   $4.95769$ \\
$10^7$ &   $5.34979$ &   $5.34771$ &   $5.34979$ &   $5.34804$ &   $5.34800$ &   $5.34768$ &   $5.34800$ &   $5.34800$ &   $5.34766$ \\
$10^8$ &   $5.40600$ &   $5.40596$ &   $5.40600$ &   $5.40597$ &   $5.40597$ &   $5.40596$ &   $5.40597$ &   $5.40597$ &   $5.40596$ \\
\end{tabular}
\tablenotetext[1]{$^{\text{-- f}}$ See the corresponding footnotes of Table \ref{table:Omega}.}
\end{ruledtabular} 
\end{table*}

\begin{figure}
\includegraphics[width=\columnwidth]{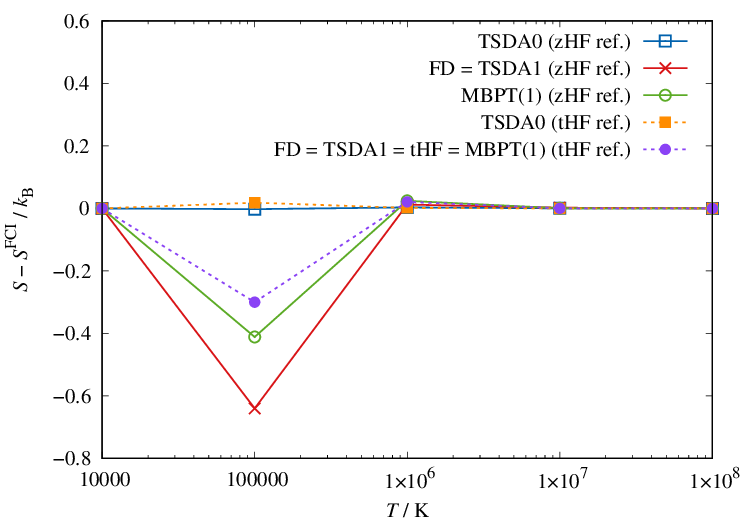}
\caption{Same as Fig.\ \ref{fig:FH_Omega} but for entropy $S$.}
\label{fig:FH_S}
\end{figure}

The entropy $S$ as a function of $T$ is given in Table \ref{table:S} and Fig.\ \ref{fig:FH_S}. Theoretically, it is expected that $S$ by any one of these methods 
is exact in both low- and high-$T$ limits. This expectation is borne out in these table and figure. The TDSA0 with either the zero-temperature or thermal HF reference
closely traces thermal FCI theory for $S$. 

\section{Conclusions}

In this article, we derived the formalisms of several {\it ab initio} thermal mean-field theories for fermions in the grand canonical ensemble. They include
well-established ones such as Fermi--Dirac theory and thermal HF theory as well as new or relatively new ones, i.e., finite-temperature MBPT(1)
and two modifications of the TSDA introduced originally by Kaplan and Argyres.\cite{Kaplan1975} 

The TSDA was proposed as an ansatz, but never implemented or developed any further. We proposed and implemented two computationally tractable modifications:
The first one (TSDA0) is the original TSDA without the stipulation of minimizing its grand potential by orbital rotation. This has been implemented into a modified thermal FCI program that 
neglects all off-diagonal elements of the FCI Hamiltonian matrix. The second modification (TSDA1) invokes the lowest-order Ursell--Mayer-like cumulant expansion of the grand partition function, leading to 
a non-converged thermal HF theory. It, therefore, hardly constitutes a new theory, but serves as an alternative derivation or justification of thermal HF theory starting from
a single, consistent grand partition function. 

The first-order perturbation corrections to the chemical potential and entropy are shown to be zero, when thermal HF theory
is used as the reference. Therefore, finite-temperature MBPT(1) offers a third way of deriving formalisms of thermal HF theory
from a grand partition function. We have shown, in some cases by explicit evaluations, that all fundamental thermodynamic relations are 
obeyed by Fermi--Dirac theory, thermal HF theory, TSDA0, and finite-temperature MBPT, confirming Argyres {\it et al.}\cite{Argyres1974} 

Numerically, the TSDA0 and finite-temperature MBPT(1) are shown to be relatively insensitive to the choice of the reference and work well, 
whereas the TSDA1 with the zero-temperature HF reference
tends to suffer from larger errors. This underscores the importance of correctly adjusting the temperatures used in the reference theory and subsequent treatment.
Overall, in spite of some irregularities in its derivation, thermal HF theory proves to be robust both analytically and numerically.

In the following article,\cite{Hirata24} thermal HF theory is extended to include the effect of electron correlation, while maintaining its quasi-independent-particle framework.

\acknowledgments
This work was supported by the U.S. Department of Energy (DoE), Office of Science, Office of Basic Energy Sciences under Grant No.\ DE-SC0006028 and also by the Center for Scalable Predictive methods for Excitations and Correlated phenomena (SPEC), which is funded by the U.S. DoE, Office of Science, Office of Basic Energy Sciences, Division of Chemical Sciences, Geosciences and Biosciences as part of the Computational Chemical Sciences (CCS) program at Pacific Northwest National Laboratory (PNNL) under FWP 70942. PNNL is a multi-program national laboratory operated by Battelle Memorial Institute for the U.S. DoE.
S.H.\ is a Guggenheim Fellow of the John Simon Guggenheim Memorial Foundation. 

\bibliography{library.bib}

\end{document}